\documentclass[10.75pt, a4paper]{article}

\usepackage{titlesec}
\titleformat{\section}
  {\bf\sffamily}
  {\thesection. }
  {5pt}
  {\MakeUppercase}
\renewcommand{\thesection}{\Roman{section}} 

\titleformat{\subsection}
  {\bf\sffamily}
  {\thesubsection. }
  {5pt}{}
  
\renewcommand{\thesubsection}{\Alph{subsection}}

\usepackage[affil-it]{authblk} 
\usepackage{etoolbox}
\usepackage{lmodern}

\usepackage{eurosym}

\usepackage[utf8]{inputenc}
\usepackage{geometry}
\usepackage[hidelinks]{hyperref}
\usepackage[citestyle=numeric,style=phys,biblabel=brackets,backend=bibtex,sorting=none,url=false,doi=false, natbib]{biblatex}
\usepackage{csquotes}
\usepackage[]{authblk} 
\usepackage{etoolbox}
\usepackage{lmodern}

\usepackage{amsmath}
\usepackage{enumerate}
\usepackage{enumitem}
\usepackage{graphicx}
\usepackage{siunitx}
\usepackage{float}
\usepackage[]{parskip} %skip=0.7em, indent=1.5em

\makeatletter
\patchcmd{\@maketitle}{\LARGE \@title}{\fontsize{16}{19.2}\selectfont\@title}{}{}
\makeatother

\usepackage[normalem]{ulem}

\usepackage{graphicx}% Include figure files
\usepackage{dcolumn}% Align table columns on decimal point
\usepackage{bm}% bold math
\usepackage{tikz}
\usetikzlibrary{math}
\usetikzlibrary{shapes.geometric, arrows}
\usetikzlibrary{positioning}
\usetikzlibrary{shapes,arrows}
\usetikzlibrary{intersections}
\usepackage{csvsimple}
\usepackage{changepage}
\usepackage[tbtags]{mathtools}
\usepackage{siunitx}
\usepackage{csquotes}

\usepackage{float}
\usepackage{subfigure}
\usepackage[
	nonumberlist, 				% keine Seitenzahlen anzeigen
	acronym,      				% ein Abk�rzungsverzeichnis erstellen
	nomain,						% no main glossary
	nopostdot,					% Kein punkt nach Beschreibung
]{glossaries}
\usepackage{glossary-superragged}
\usepackage[hidelinks]{hyperref}
\usepackage{color, colortbl}

\usepackage{wrapfig}

\usepackage{pgfplots}
\usepackage{tikz}
\usetikzlibrary{arrows}
\usepackage{amsmath}
\pgfplotsset{compat=newest}
\usepgfplotslibrary{fillbetween}
\usetikzlibrary{calc}
\def\centerarc[#1](#2)(#3:#4:#5)% Syntax: [draw options] (center) (initial angle:final angle:radius)
{ \draw[#1] ($(#2)+({#5*cos(#3)},{#5*sin(#3)})$) arc (#3:#4:#5); }

\usepackage{amssymb}

\usepackage{nicefrac}

\usepackage{abstract}
\usepackage{multirow}
\usepackage{tabularx}
\usepackage{booktabs}
\usepackage{array}
\usepackage{longtable}
\newcolumntype{L}[1]{>{\raggedright\let\newline\\\arraybackslash\hspace{0pt}}m{#1}}
\newcolumntype{C}[1]{>{\centering\let\newline\\\arraybackslash\hspace{0pt}}m{#1}}
\newcolumntype{R}[1]{>{\raggedleft\let\newline\\\arraybackslash\hspace{0pt}}m{#1}}

\usepackage[framemethod=tikz]{mdframed}
\usepackage{lipsum}

\usepackage{dirtytalk}
\usepackage{tcolorbox}
\usepackage[table]{xcolor}
\usepackage{caption}
\newtcolorbox{mybox}[1]{colback=green!6!white,colframe=black!75!black,fonttitle=\bfseries,title=#1}
\newtcolorbox{mybox2}{colback=red!5!white,colframe=red!75!black}

\usepackage{pifont}

\def\be{\begin{equation}}
\def\ee{\end{equation}}

\usepackage{soul,xcolor}
\setstcolor{red}

\usepackage{xcolor,hyperref}
\hypersetup{
   colorlinks,
   linkcolor={blue!50!black},%{red!80!black},
   citecolor={blue!50!black},
   urlcolor={blue!80!black}
} 
\definecolor{mycolor}{rgb}{0.122, 0.435, 0.698}

\definecolor{revblue}{rgb}{0.0, 0.0, 1.0}
\definecolor{revpink}{rgb}{0.90, 0.10, 0.55}

\title{Elasto-Plastocapillarity:\\
Surface-Tension-Driven Flows of Yield-Stress Fluids}
\author[1,2]{Maziyar Jalaal\footnote{m.jalaal@uva.nl, m.jalaal@damtp.cam.ac.uk, ORCID: 0000-0002-5654-8505}}

\affil[1]{Van der Waals-Zeeman Institute, Institute of Physics, \protect\\
University of Amsterdam, Science Park 904, Amsterdam, 1098XH, The Netherlands}
\affil[2]{Department of Applied Mathematics and Theoretical Physics, \protect\\
University of Cambridge, Wilberforce Road, Cambridge CB3 0WA, UK}

\begin{document}
\begingroup
\sffamily
\date{}
\maketitle
\endgroup

\begin{abstract}
In this perspective article, we introduce \emph{elasto-plastocapillarity}: surface-tension-driven free-surface flows of yield-stress fluids in which capillary stresses compete with yielding and elastic relaxation. We frame the relevant parameter space, and highlight representative problems (drops, bubbles, jets/filaments, thin films, and porous media) that are central to applications ranging from additive manufacturing and coatings to biological flows. 
We then outline a set of timely open challenges, including surface-tension measurements, breakdown of scale separation near the microstructures, coupled bulk-interface flow experimental diagnosis, interfacial rheology and surface stresses, history dependence, constitutive models and boundary conditions, reduced theoretical models, and the emerging opportunity provided by artificial intelligence and microgravity; aimed at guiding future experiments, theory, and computation.

\textbf{keywords: Yield Stress Fluids $|$ Plastocapillarity $|$ Elasto-Viscoplasticity $|$ Surface Tension} 

\end{abstract}

%%%%%

%%%%%%%%%%%%%%%%%%%%%%%%%%%%%%% Introduction %%%%%%%%%%%%%%%%%%%%%%%%%%%%%%%%%%%%%%%%%%%%%%
% \textcolor{red}{TO DO:}\\
% Anniina's work on bubbles and diffusion in YSF \\   -> done (intro, foam coarsening)
% Ref 16 and 76 are the same\\                        -> done (also 12/83, 13/38, 25/87 were duplicates)
% Emad: new wet porous media work\\                   -> done (intro, porous media)
% Rev1: make a better clarification on what is inherent to new method developments?\\  -> done (pink)
% Rev1: Memory: The link between thermodynamics and “history dependent” dynamical effects.\\  -> done (new subsection)
% Rev2: more concrete examples (experiments in particular), that would illustrate the theoretical arguments\\  -> done (end of Sec. I + experiments)
% Rev2: some references given the relatively short text. Ref suggest to shorten, but maybe I should expand.\\  -> see chat
% Gareth: L, e.g., the diameter of ...\\              -> done
% Gareth: Enhanced Oil recovery for wet porous media\\  -> done
% Gareth: microfluidic paper Thibaut sent for measuring small yield stress - maybe next Stefano's\\  -> done (experiments)
% Gareth and Mazi: when talk about moels talk about pre-streess and new problems raised by Neil et al - and say KIH may be immune to those problems\\  -> done (computational methods)
% Gareth: spelling mistake in figure 2 -> fixed (Figures/Fig2-v2.png)

\section*{What Is Elasto-Plastocapillarity? Why Does It Matter?}
Yield-stress fluids form a broad class of soft materials that behave like solids at low stress but flow like liquids once a finite yield stress $\tau_0$ is exceeded. The  microstructural origins of this response vary widely, spanning jammed colloidal suspensions and emulsions/foams to polymeric and particulate gels; yet, at the macroscopic (continuum) level, many such materials share an \emph{elasto-viscoplastic (EVP)} character~\cite{bonn2017yield}. 

In many practical settings, yield-stress fluids flow with a free surface. Classical large-scale examples occur in geophysical contexts (\emph{e.g.}, mudslides, debris flows, lava, and glaciers), where gravity dominates and surface tension is typically negligible~\cite{ancey2007plasticity,balmforth2025implications}. Capillary forces become relevant only when the characteristic length scale of the problem, $\mathcal{L}$ (\emph{e.g.}, the radius of a bubble or droplet or the thickness of a film), is comparable to or smaller than the capillary length $
\Lambda=\left(\sigma/\rho\,g\right)^{1/2}$,
where $\sigma$ is the surface-tension coefficient, $\rho$ the density, and $g$ gravity~\cite{gennes2004capillarity}. This small-scale limit is precisely where \emph{Elasto-Plastocapillarity} becomes central, underpinning many modern applications including additive manufacturing and bioprinting (\emph{e.g.}, direct ink writing, embedded printing), thin-film coating and lubrication, and food processing. Related problems, such as jets~\cite{balmforth2010extensional,shemilt2025rayleigh}, filaments~\cite{van2023viscoplastic,hossain2025critical,sauret2026fluid,franca2026coalescence}, thin films~\cite{shemilt2025air},  droplet spreading~\cite{luu2009drop,jalaal2021spreading,martouzet2021dynamic,francca2024elasto,mitra2024rapid}, bubble rise and growth~\cite{pourzahedi2022flow,esposito2025rising,balasubramanian2024bursting,requier2024foam}, often display nontrivial dynamics due to the interplay of elasticity, yielding, and capillarity.
Capillary-driven transport in porous media~\cite{singh2019capillary,chaparian2020yield,chaparian2025comprehensive} is another key example, with important implications for filtration, medical devices and enhanced oil recovery~\cite{lake2014fundamentals}. Small-scale free-surface flows also arise in biological systems, such as mucus in airways and the gut~\cite{bansil2013influence,kavishvar2023yielding,shemilt2025viscoplasticity}, where rheology can be a decisive factor in health and disease, and in ecological contexts, such as microbial colonies confined in yield-stress matrices~\cite{vajdi2026biogenic}. Finally, in microgravity, surface tension dominates even at large length scales, a regime relevant to space science and engineering~\cite{van20233d,heitmeier2025spreading}.

\begin{figure}[t!]
\centering
\includegraphics[width=0.85\textwidth]{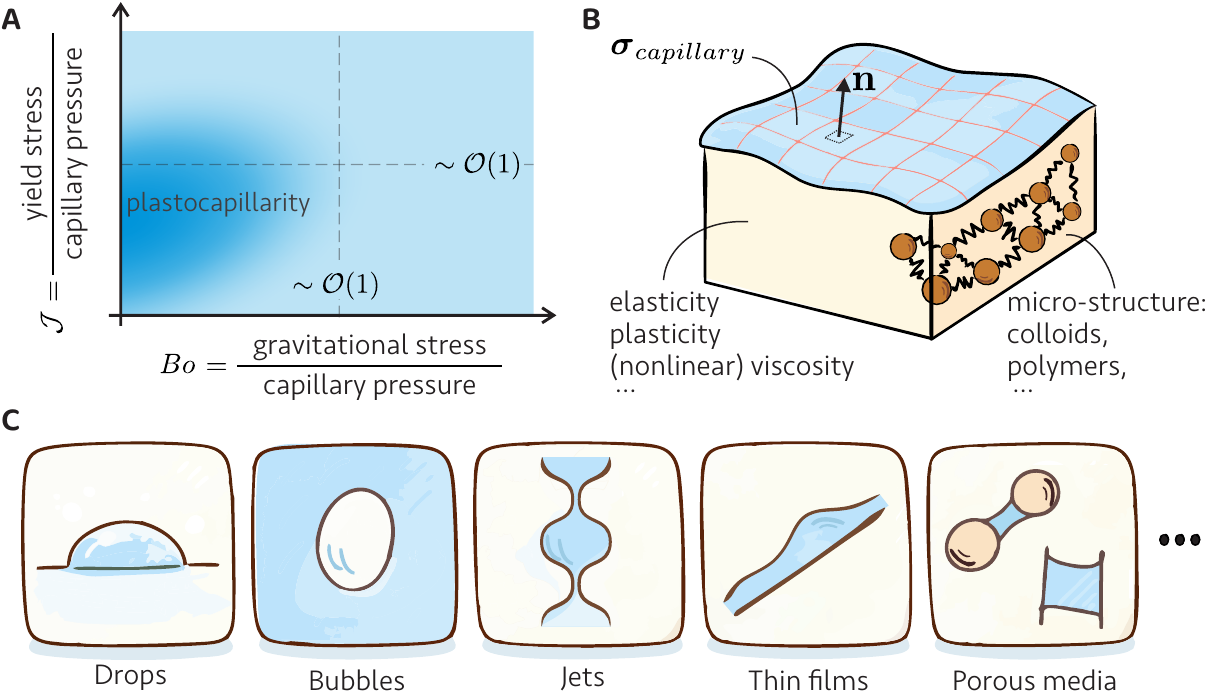}
\caption{\textbf{A.} Schematic regime map for free-surface flows of yield-stress fluids in terms of the plastocapillary number $\mathcal{J}=\tau_0 \mathcal{L}/\sigma$ (yield stress relative to capillary pressure) and the Bond number $Bo=\rho g \mathcal{L}^2/\sigma$ (gravity relative to capillarity). The shaded region (darker blue) highlights the plastocapillary window.
\textbf{B.} Conceptual sketch of a yield-stress material with an underlying microstructure (colloids, polymers, \dots) and a deformable free surface; capillary traction acts along the interface with a normal vector $\mathbf{n}$, while the bulk response combines solid-like and liquid-like behaviour.
\textbf{C.} Representative settings governed by elasto-plastocapillarity: drops, bubbles, jets/filaments, thin films, and porous-media imbibition.}
\label{fig1}
\end{figure}

In the \emph{Plastocapillarity} limit~\cite{jalaal2023plastocapillarity} (neglecting elasticity and other rheological details), two dimensionless groups are particularly useful:
\begin{equation}
Bo = \frac{\rho\,g\,\mathcal{L}^2}{\sigma},
\quad \mathrm{and} \quad
\mathcal{J}=\frac{\tau_0\,\mathcal{L}}{\sigma},
\end{equation}
where the Bond number ($Bo$) compares gravitational to capillary stresses and the plastocapillary number ($\mathcal{J}$) compares the yield stress to the characteristic capillary stress $\sigma/\mathcal{L}$. When $Bo\gg 1$ (\emph{i.e.}\ $\mathcal{L}\gg \Lambda$), gravity overwhelms capillarity; when $Bo\lesssim 1$, capillary effects are prominent. Within this capillary-dominated range, plastocapillary behaviour is strongest when $\mathcal{J} <O(1)$: capillary pressures are large enough to deform the material, yet yielding still constrains the flow. For $\mathcal{J}\gg 1$, capillary stresses are too weak to significantly deform or mobilise the solid-like material, so plastocapillary effects become weak (see Fig.~\ref{fig1}A). Note that other groups (\emph{e.g.}, the Ohnesorge number, which measures viscous dissipation) also matter, but we largely omit them in this short note for clarity.

Because yield-stress fluids are inherently elastic, at least one additional parameter is needed to characterise such effects. A natural choice is the Deborah number
\begin{equation}
De = \lambda \, \left( \frac{\sigma}{\rho\,\mathcal{L}^3}\right)^{1/2},
\end{equation}
\emph{i.e.}\ the ratio of the elastic relaxation time $\lambda$ to the capillary time scale.
Alternatively, depending on the chosen nondimensionalisation, one may characterise the same physics using an \emph{elastocapillary} measure, \emph{e.g.} comparing elastic to capillary stresses through $Ec = G\,\mathcal{L}/\sigma$, where $G$ is a representative elastic modulus; such forms are common in elastocapillarity~\cite{mckinley2005, andreotti2020statics,style2017elastocapillarity,bico2018elastocapillarity} and can be mapped to $De$ for a given constitutive relation.

While the systematic role of elasticity across plastocapillary regimes and for different geometries 
and applications remains incompletely mapped, it is expected to strongly influence problems such as 
jet breakup, droplet spreading, and filament thinning. It is worth noting that 
elasto-plastocapillarity connects naturally to two well-established limiting regimes of elasto-capillarity 
in soft solids and complex fluids. When the 
applied stresses are small compared to the yield stress ($\mathcal{J} \gg 1$), the material behaves 
predominantly as a viscoelastic solid, and the classical regime of 
\emph{solid elastocapillarity} is asymptotically approached \cite{style2017elastocapillarity, bico2018elastocapillarity}. In this limit, the relevant phenomena include the deformation 
of a soft elastic substrate by a sessile drop, forming the elastocapillary wetting ridge, 
modification of contact angles away from Young's law, \emph{etc}. In the opposite limit of a 
viscoelastic liquid with a small or negligible yield stress ($\mathcal{J} \ll 1$, $De > 0$), the 
regime of \emph{visco-elasto-capillarity} is recovered, where elastic stresses compete with 
capillarity during free-surface deformation. Here the paradigmatic phenomena are capillary thinning 
and breakup of viscoelastic filaments, which develop the characteristic beads-on-a-string morphology 
and exhibit elastocapillary thinning regimes governed by the balance of polymer relaxation stresses 
against surface tension \cite{mckinley2005, clasen2006beads,gaillard2025does}. Elasto-plastocapillarity thus occupies the 
rich intermediate territory between these two limits, where yielding, elasticity, and capillarity 
are simultaneously relevant.

\begin{figure}[t!]
\centering
\includegraphics[width=0.75\textwidth]{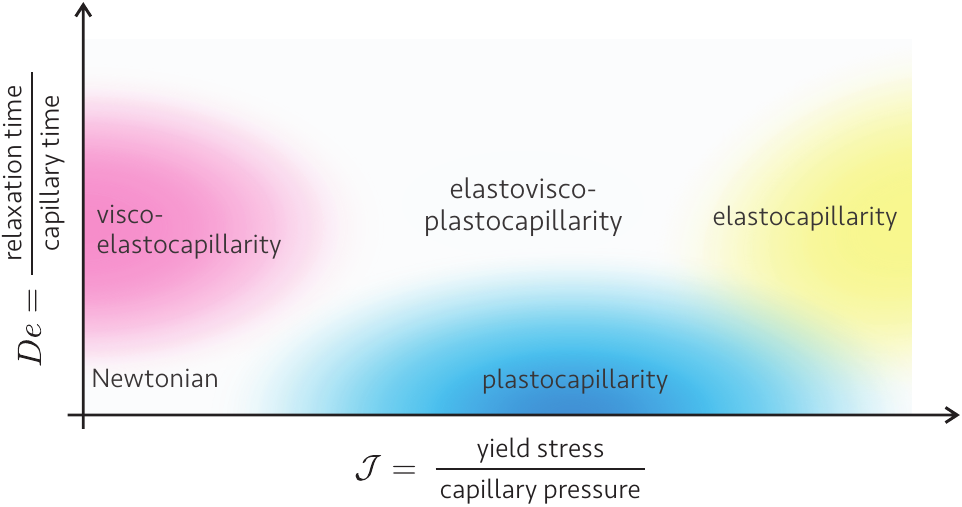}
\caption{Regime map for capillary-driven free-surface flows of yield stress fluids in the 
$(\mathcal{J}, De)$ plane, where $\mathcal{J}$ is the plastocapillary 
number (yield stress relative to capillary pressure) and $De$ 
is the Deborah number (elastic relaxation time relative to capillary time), evaluated at 
fixed polymer viscosity.}
\label{fig2}
\end{figure}

This intermediate territory, and its connections to the limiting regimes, can be 
visualised systematically in the $({\mathcal J}, De)$ plane, as shown in Fig.~\ref{fig2}. 
For a fixed polymer viscosity, varying $\mathcal{J}$ and $De$ independently recovers all relevant
limits. In the bottom-left corner ($\mathcal{J} \to 0$, $De \to 0$), the fluid is essentially
Newtonian: capillarity drives flow freely, without resistance from either yielding or elastic relaxation. Moving upward along
the $De$ axis at small $\mathcal{J}$, the elastic relaxation time grows relative to the capillary time scale
and the visco-elasto-capillary regime is entered, manifested by filament thinning or
beads-on-a-string. Moving rightward along the $\mathcal{J}$
axis at small $De$, yielding dominates and the plastocapillary regime is recovered, where
capillary-driven flow arrests once stresses fall below the yield stress.  At large $\mathcal{J}$ and
finite $De$, the material behaves increasingly as a viscoelastic solid and the soft-solid elastocapillary
regime is asymptotically approached. Finally, the central region of
the diagram defines the core of
elasto-viscoplastocapillarity: here yielding, elastic relaxation, and capillarity are
simultaneously relevant. Existing
results from Newtonian capillarity, visco-elasto-capillarity, and soft-solid elastocapillarity each
occupy well-defined regimes of this diagram, while the largely uncharted interior is the
territory that elasto-plastocapillarity seeks to explore.
Canonical free-surface problems are beginning to populate this interior: droplet spreading (where does a drop stop?) and bubble rise (when does a bubble start to move?) are now backed by substantial theory and well-controlled experiments~\cite{jalaal2021spreading,francca2024elasto,pourzahedi2022flow,esposito2025rising}. The $(\mathcal{J},De)$ plane of bubble bursting has been mapped numerically, revealing four regimes and a non-monotonic effect of elastic relaxation on the Worthington jet~\cite{balasubramanian2024bursting}. For filaments, experiments give a critical plastocapillary number ($\approx 0.2$) for the breakup of an embedded Newtonian thread~\cite{hossain2025critical} and a critical yield stress beyond which printed filaments no longer coalesce~\cite{franca2026coalescence}, while theory shows that a yield stress suppresses the Rayleigh--Plateau instability, that sub-yield elasticity re-admits it above a critical Weissenberg number~\cite{shemilt2025rayleigh}, and that elastic stresses control the self-similar thinning near pinch-off~\cite{zakeri2025scaling}. 
Despite this progress, a complete picture that closes the gap between experiments, theory, and simulations is still missing in most of these cases. Moreover, many other problems remain essentially unexplored. Different types of interfacial instabilities in EVP materials, for instance, are of considerable importance in coating and printing problems or medical applications~\cite{shemilt2025air,ShemiltEtAl2022TubeFilm}, but have so far received little attention. Such problems are, in our view, what makes the technical effort worthwhile. In what follows, we highlight a
selection of overarching challenges and open questions, without attempting an exhaustive review of
the existing literature.

% While the systematic role of elasticity across plastocapillary regimes and for different geometris and applicaitons remains incompletely mapped,
% existing evidence shows that elasticity can be essential, for example, to recover correct bubble shapes in yield-stress fluids~\cite{}, or  and 
% it is expected to strongly influence problems such as jet breakup, droplet spreading, and filament thinning. This motivates an effort to chart elasto-plastocapillary regimes through experiments, theory, and computation, and to identify the governing mechanisms, scaling laws, and robust dimensionless phase maps. In what follows in this short perspective, we highlight a selection of overarching challenges and open questions, without attempting an exhaustive review of the existing literature.

\section*{Open Questions and Challenges}

\subsection*{(i) Scientific questions}

\subsection*{Surface tension of yield-stress fluids and its measurement.} Reliable values of the liquid--vapour surface tension coefficient are fundamental to elasto-plastocapillarity but remain surprisingly hard to establish for yield-stress materials. Reported values for the canonical model system Carbopol span a noticeable range across the literature, partly because many standard tensiometry protocols (pendant drop, capillary bridges, \emph{etc.}) do not take into account irreversible plastic deformations. As a result, classic methods can show strongly history-dependent apparent values of surface tension that are coupled to elasto-plasticity~\cite{BoujlelCoussot2013SurfaceTensionYSF,Jorgensen2015SurfaceTensionBridge}. In capillary-bridge experiments on Carbopol, for example, the \say{apparent} surface tension measured after stretching the bridge is consistently higher than after compressing it~\cite{Jorgensen2015SurfaceTensionBridge}. Newer approaches such as needle-induced cavitation rheology~\cite{MohammadigoushkiShoele2023CavitationRheology} offer an alternative route and report near-water values for Carbopol across a range of yield stresses, but systematic cross-calibration between methods and across material classes beyond Carbopol (emulsions, foams, colloidal gels) is still lacking. More fundamentally, recent simulations of glass-forming liquids~\cite{HeitmeierVoigtmann2025ShearRateSurfaceTension} indicate further complexities, where the surface tension inferred from stress anisotropy can become shear-rate dependent, because bulk normal stresses contaminate the interfacial stress balance. Determining when the surface tension coefficient of yield-stress fluids is a true thermodynamic parameter, and when it is an emergent out-of-equilibrium quantity tied to rheology and history, is therefore a core open problem for elasto-plastocapillarity.

%%%%%

\subsection*{Small-scale phenomena and breakdown of scale separation.} Elasto-plastocapillarity typically assumes a continuum description with a clear separation between the microstructural length scale (\emph{e.g.,} droplet size in emulsions, bubble size in foams, network mesh size in gels) and the macroscopic scales setting interface curvature. Yet, many practical flows (notably contact-line motion of droplets on a surface, thin coatings, and microfluidic confinement) feature geometrical length scales comparable to the size of the microstructural units. In such regimes, the continuum assumption is invalidated; hence, such problems
require corrections to theories and computations, but also, importantly, experimental techniques able to pin down the importance of such effects.
Closely related is the interaction of yield-stress fluids with solid surfaces, even far from free interfaces, where ``apparent slip'' can significantly modify the fluid mechanics~\cite{JalaalBalmforthStoeber2015SlipSpreading,martouzet2021dynamic,chaparian2021sliding}. In capillary imbibition, for instance, yield-stress fluids keep creeping as the rise slows down, at stresses nominally below the yield stress, a behaviour that is captured only once wall slip is included alongside the bulk rheology~\cite{KimChoi2024CapillaryImbibition}. Therefore, a coherent multiscale framework (linking microstructure, wall/interfacial boundary conditions, and macroscopic capillary driving) remains largely missing.

%%%%%%%%%

\subsection*{Surface stresses and interfacial rheology.} Many yield-stress interfaces are not \say{clean}: surfactants, polymers, proteins, and even adsorbed particles or macromolecules can endow the interface with its own rheology (surface viscosity/elasticity) and generate Marangoni stresses through interfacial transport~\cite{fuller2012complex}. Such \emph{surface stresses} can be decisive, because the relevant interfacial degrees of freedom may relax on fast time scales comparable to those probed by capillary and bubble-driven deformations in complex fluids, and can therefore contaminate the interpretation of an effective, constant surface tension coefficient and strongly affect the dynamics~\cite{Garbin2026BubbleDynamicsComplexFluids,Kloek2001BubbleDissolutionInterfacialRheology}. For instance, for yield-stress films in tubes (a model relevant to airway closure and coating flows), reduced-order models show that surfactant can rescale instability time scales and amplify yield-stress effects, shifting plug-formation thresholds and delaying collapse \cite{ShemiltEtAl2022TubeFilm,ShemiltEtAl2023SurfactantTubeFilm}. A pressing open question is hence to determine the limits of regime maps in which (i) bulk plasticity arrests deformation, (ii) Marangoni stresses immobilise the interface, or (iii) surface viscoelasticity or elasto-viscoplasticity~\cite{alicke2025yielding} introduces new time scales and failure modes (delayed rupture, stick--slip, \emph{etc}), with direct implications for plug formation, film rupture, and print/coating outcomes. Addressing this will require improved metrology of interfacial properties and new benchmark experiments/simulations that couple EVP bulk rheology with interfacial transport and surface rheology in a controlled way.

%%%%%%%%%

\subsection*{History dependence, memory, and residual stresses.}

Capillary-driven flow in a yield-stress material stops as soon as the stress falls below the yield stress everywhere in the material, not when the surface energy is minimised. The arrested shape may therefore not be unique and depend on the path (history) taken to reach it. Because plastic deformation is irreversible, there is no restoring force to erase this mechanical memory once the driving stops, so the shape of a deposited drop, a printed filament, or a coated film depends on how it was made. Experiments on Carbopol drops indeed show that spreading arrests at a mechanical, rather than thermodynamic, equilibrium, with a final contact angle set by the yield stress, the drop size, and the boundary condition at the wall~\cite{jalaal2021spreading,martouzet2021dynamic}. 
For $\mathcal{J}\ll1$, capillary stresses are strong enough to yield the material everywhere, and any the early memory of deformation is largely erased; as $\mathcal{J}$ grows towards $O(1)$, capillarity can no longer fully yield the material, parts of it never flow, and the initial condition and deformation history become increasingly imprinted on the final state. In this regime, the arrested shape of a spreading droplet depends on the deposition protocol, such as the position and speed of the nozzle, its distance from the substrate, and the extrusion rate. 
Any surface tension inferred from such shapes is then entangled with the mechanical history of the sample, which is one reason why the thermodynamic status of $\sigma$ (see above) is so hard to pin down. Elasticity adds further layers. For instance residual stresses left by the preparation remain locked in the unyielded regions, set the initial stress state (pre-stress) that enters the constitutive models discussed below, and may be released later upon ageing or re-loading, thereby altering the subsequent dynamics. Making the link between the history-dependent states and their dynamics, and designing experiments in which the preparation protocol is varied systematically, is a promising open directions in the field.

\subsection*{(ii) Methodological challenges}

\subsection*{Experiments: beyond interface shapes.} Much of the existing experimental literature infers mechanics from interface profiles (\emph{e.g.,} bubble and drop shapes). Recent microfluidic approaches have pushed this 
further by extracting fast oscillatory rheology from the time-dependent deformation of 
transported droplets in shape-optimised channels \cite{milani2026microfluidic}. Extending such 
techniques to yield-stress droplets would offer a direct handle on the interplay 
between bulk plasticity, elasticity, and interfacial response under controlled 
capillary forcing. Further progress now hinges on methods that can resolve both bulk kinematics/stress characteristics and interface geometry and dynamics at matched spatio-temporal resolution. Classic methods such as particle image velocimetry (PIV) and micro-PIV have already shown great potential for flows near gas--liquid interfaces~\cite{holenberg2013ptv,MohammadigoushkiShoele2023CavitationRheology} or solid boundaries~\cite{KimChoi2024CapillaryImbibition,nazari2023helical}. More recent exciting developments include methods to directly reconstruct stress fields using high-speed photoelastic tomography~\cite{yokoyama2024high,igarashi2025reconstruction}; however, such methods still need to be tested for complex systems with deformable interfaces and complex geometries. Therefore, simultaneous measurements of bulk flow and interfaces in 3D at high spatio-temporal resolution remain a challenge. Optical coherence tomography (OCT)~\cite{jalaal2018gel,daneshi2019characterising,van2023viscoplastic, amini2025optical} is particularly attractive for small-volume flows, as it can capture cross-sectional free-surface profiles in various geometries and, when combined with Doppler processing or microrheology-like methods, access velocity fields in opaque, structured fluids \cite{jalaal2018gel,AminiEtAl2024DOCTTEVP}. Where optical access is limited, as in many biological problems and industrial applications, ultrasound-based velocimetry provides a complementary \emph{in-situ} route to internal flows and rheological inference~\cite{GurungEtAl2016UIVRheology}. Translating these diagnostics to \emph{in-vivo} settings is challenging but increasingly plausible~\cite{OldenburgEtAl2012OCTMucus,ChuEtAl2016MicroOCT}. At even smaller scales (\emph{e.g.,} condensates inside a cell), combining microscopy with active microrheology may be essential for accessing nonlinear rheology and interfacial dynamics simultaneously~\cite{Jawerth2020ProteinCondensatesMaxwell,SquiresBrady2005ActiveMicrorheology,SquiresMason2010Microrheology,PuertasVoigtmann2014MicrorheologyColloids}.

%%%%%%%

\subsection*{Computational Methods.} The coupled problem of elasto-viscoplastic bulk rheology and free-surface dynamics has attracted growing attention, and recent work has begun to map how yielding, elastic relaxation and capillary stresses interact across droplets, bubbles, filaments and films. A common route is to combine an EVP constitutive law (often the Saramito model~\cite{saramito2009new}) with an interface-capturing method such as Volume-of-Fluid (VOF)~\cite{HirtNichols1981VOF,popinet2018numerical}, which yields a comparatively simple and computationally tractable framework for strongly deforming interfaces~\cite{francca2024elasto,zakeri2025scaling,esposito2025rising}. Simulations, however, remain limited by several well-known bottlenecks: the accurate localisation of yield surfaces; sensitivity to regularisation choices used to handle the yield criterion numerically~\cite{Papanastasiou1987YieldRegularization,frigaard2005usage,saramito2017progress}; and the fact that many real yield-stress materials exhibit additional physics (\emph{e.g.}, thixotropy, ageing, pre-yielding nonlinearities, and nonlocality) not captured by standard EVP closures~\cite{FraggedakisDimakopoulosTsamopoulos2016EVPComparison,MewisWagner2009Thixotropy,SerialEtAl2021NonlocalThixotropic,kamani2021unification,divoux2024ductile}. A further subtlety, highlighted recently by Shemilt, Balmforth \& Hewitt~\cite{shemilt2026startup}, concerns the initial stress state, or \emph{pre-stress}, in Saramito-type models: because the sub-yield material is an elastic solid, the flow can depend on the stress configuration prior to loading (base states are non-unique), and stress discontinuities may develop at yield surfaces unless a suitable pre-stress is prescribed. The consequences of this for capillary flows remain unclear and deserve systematic study (see also the discussion of history dependence above). Constitutive frameworks rooted in plasticity theory with kinematic hardening, such as the isotropic--kinematic-hardening (IKH) family of models~\cite{dimitriou2014comprehensive,dimitriou2019canonical}, in which yielding is governed by the stress relative to an evolving back stress and no sharp flow/no-flow switch is required, may be immune to some of these complexities, but this remains to be studied. Alternatives to regularised methods are provided by augmented-Lagrangian/variational-inequality methods, which can deliver highly accurate stress fields and yield-surface locations in viscoplastic settings~\cite{RoquetSaramito2003BinghamCylinder,Zhang2011AugLagBubble,pourzahedi2022flow,balmforth2025implications,chaparian2025comprehensive}. These methods are typically more expensive and can be difficult to extend to fully time-dependent, strongly deforming interfaces, but they provide valuable benchmarks for validating regularised approaches and theories. Finally, boundary conditions are often decisive: wall slip is pervasive in yield-stress fluids and can qualitatively change free-surface evolution, yet robust, transferable slip/contact-line laws for EVP materials and implementation in computational codes are still lacking~\cite{Barnes1995WallSlipReview,BalmforthFrigaardOvarlez2014ViscoplasticReview,chaparian2021sliding}.

%%%%%
\subsection*{Reduced Theoretical Models.} Existing thin-film and long-wave theories capture the dynamics of shallow gravity-driven flows of viscoplastic fluids~\cite{balmforth1999consistent,balmforth2002shallow}, and similar ideas have informed recent models for capillary-driven viscoplastic droplet spreading and thin films~\cite{jalaal2021spreading}. However, a systematic reduced-order theory for elastoviscoplastic capillary flows is still lacking. Such models are important because they can yield simple criteria and scaling laws for dynamics, stability, and arrest of films and droplets, with direct relevance across many applications. In this context, further analysis of EVP constitutive models and their asymptotic limits~\cite{Saramito2007EVPConstitutive,FraggedakisDimakopoulosTsamopoulos2016EVPComparison,kamani2021unification} remains a crucial step. Beyond this, additional challenges include incorporating consistent contact-line laws, spatially varying surface properties, non-isothermal effects, and realistic wall interactions (including slip) within a unified thin-film EVP framework (see notes above).

% Most current free-surface simulations combine an EVP constitutive equation (often of Saramito type) with an interface-capturing method (e.g. VOF), providing tractable frameworks for elastoplastocapillary computations. \cite{Saramito2007EVPConstitutive,HirtNichols1981VOF,FrancaJalaalOishi2024EVPSpreading} Predictive capability is still constrained by (a) the difficulty of resolving sharp yield surfaces and unyielded plugs, (b) sensitivity to regularisation and discretisation choices, and (c) incomplete constitutive structure for many real materials (ageing, thixotropy, nonlocality). Augmented-Lagrangian and variational-inequality methods can provide accurate stress fields and yield-surface locations in viscoplastic settings, but are more challenging to extend to strongly deforming interfaces. \cite{Zhang2012AugLagBubble,RoquetSaramito2003BinghamCylinder,Papanastasiou1987YieldRegularization,FraggedakisDimakopoulosTsamopoulos2016EVPComparison} Equally decisive are boundary conditions: wall slip is pervasive in yield-stress materials and can control capillary rise and arrest, while contact-line laws for EVP/YSF remain far from settled. \cite{Barnes1995WallSlipReview,KimChoi2024CapillaryImbibition,JalaalBalmforthStoeber2015SlipSpreading,Coussot2014YieldStressReview,BalmforthFrigaardOvarlez2014ViscoplasticReview,MewisWagner2009Thixotropy,Serial2021NonlocalThixotropy}

%%%%%%

\subsection*{Microgravity: a new experimental platform for elasto-plastocapillarity.} Microgravity is increasingly relevant for space manufacturing and life-science technologies, where capillary forces dominate fluid handling and interface control. It also provides a distinctive opportunity for fundamental studies: reducing hydrostatic stresses allows much larger scales to remain in the capillary (small-$Bo$) regime, facilitating diagnostics and improving signal-to-noise for interface measurements~\cite{DAngeloEtAl2022VIPDROP2}. Recent drop-tower experiments on yield-stress droplet spreading explicitly exploit this to disentangle Bond and plastocapillary effects and to probe parameter ranges inaccessible on Earth~\cite{heitmeier2025spreading}. The main challenges remain platform-specific (short time windows, g-jitter, vibration, remote operation and constrained sample preparation), but mature infrastructure exists across drop towers, parabolic flights and sounding rockets, offering a realistic pathway to reproducible benchmark data sets for the community.

\subsection*{AI \& Data-driven methods for inference.}

Elasto-plastocapillary experiments naturally generate rich data streams (interface contours, curvature fields, and bulk velocimetry) that, when combined with robust theory, can be exploited for \emph{inverse rheology}: inferring rheological parameters (yield stress, elastic moduli, shear-thinning indices, \emph{etc.}) as well as effective surface-tension coefficients, and even interfacial transport parameters. 
Physics-informed neural networks (PINNs) and related frameworks enable such inference by embedding the governing PDEs and constitutive relations as constraints~\cite{Raissi2019PINNs,Karniadakis2021PIML}.
\emph{Rheology-informed} neural architectures for forward prediction and inverse parameter identification of complex-fluid constitutive models~\cite{MahmoudabadbozchelouJamali2021RhINNs}, their extension to non-Newtonian flow fields, including the recovery of unknown wall-slip conditions from sparse velocity data~\cite{MahmoudabadbozchelouKarniadakisJamali2022nnPINNs}, and practical guidelines for parameter recovery~\cite{SaadatMangalJamali2023Guideline} provide another natural route to tackle elasto-plastocapillarity problems. 
Similarly, related efforts toward \emph{digital rheometer twins}~\cite{MahmoudabadbozchelouEtAl2022DigitalRheometerTwins} and frame-invariant, learnable constitutive models that are trained on rheometric data and deployed directly in multidimensional simulations~\cite{lennon2023scientific} offer practical templates for learning the material response from a limited set of rheometric protocols and transferring it to complex flows. Neural operators offer yet another alternative, and have recently shown success in learning the response of complex fluids to arbitrary flow protocols~\cite{mangal2025learning}.
Complementary to neural approaches, \emph{Bayesian inference} provides a way to quantify uncertainty and \emph{model discrepancy} when transferring parameters from simple rheometry to complex flows~\cite{rinkens2026bayesian}, an approach that could be valuable for elasto-plastocapillary flows, where arrest, onset, and breakup depend sensitively, through $\mathcal{J}$, on the values of yield stress and surface tension adopted. \\
% For instance, recent Bayesian model selection for Carbopol shows that rheology-informed priors can fail in squeeze flows, while broader (expert-informed) priors recover predictive posteriors and clearer model plausibilities \cite{rinkens2026bayesian}.
% Other Additionally, operator-learning methods offer a route to accelerate multiphase EVP forward solves inside Bayesian/optimization loops, potentially making inference feasible at the level of full free-surface simulations \cite{LuEtAl2021DeepONet,LiEtAl2020FNO}.

In summary, elasto-plastocapillarity offers a fertile playground where capillarity,
yielding, and relaxation co-create new interfacial phenomena and arrest mechanisms
across a wide range of scales. The field sits at a remarkable crossroads: it can be 
approached from the perspective of solid mechanics, where capillary forces deform soft structured materials, or from fluid mechanics, where free-surface flows 
are arrested or reshaped by rheological complexity. At the smallest 
scales, it is deeply connected to physical chemistry, from the thermodynamic status 
of surface tension in out-of-equilibrium soft matter, to the adsorption and transport 
of surfactants, polymers, and particles at deformable interfaces. At intermediate 
scales, microstructural physics governs how colloidal suspensions, emulsions, foams, 
and gels yield, relax, and age in the presence of capillary stresses. At larger scales 
still, the same competition between capillarity and rheological resistance shapes 
phenomena as diverse as mucus transport in airways and the printing of 
functional inks and bioinks. The field is therefore wide open to theorists, 
computationalists, and experimentalists alike, drawing on expertise from soft matter 
physics, chemical engineering, applied mathematics, biophysics, and geophysics, with 
cross-cutting impact in additive manufacturing, coating and filtration, biomedical 
engineering, and earth sciences. Establishing robust measurements, multiscale models, 
and benchmark experiments across this breadth is a natural next step.

\section*{Acknowledgements}
The author thanks Hugo Fran\c{c}a, Alice Woodbridge, Duncan Hewitt, Safa Jamali, Emad Chaparian, and Gareth McKinley for comments on the manuscript, and acknowledges support from the Dutch Research Council (NWO) through the Vidi project Living Levers (No.\ 279.21.239) and from the European Research Council (ERC) under the European Union's Horizon Europe research and innovation programme through the Starting Grant FluMAB (grant agreement No.\ 101117025).

% %%%%%%%%%%%%%%%%%%%%%%%%%%%%%%% Competing interests %%%%%%%%%%%%%%%%%%%%%%%%%%%%%%%%%%%%%%%%%%%%%%
% \section*{Competing interests}
% The authors declare no competing interests.

% %%%%%%%%%%%%%%%%%%%%%%%%%%%%%%% Author contributions %%%%%%%%%%%%%%%%%%%%%%%%%%%%%%%%%%%%%%%%%%%%%%
% \section*{Author contributions}

%%%%%%%%%%%%%%%%%%%%%%%%%%%%%%% Funding %%%%%%%%%%%%%%%%%%%%%%%%%%%%%%%%%%%%%%%%%%%%%%
% \section*{Funding}

%%%%%%%%%%%%%%%%%%%%%%%%%%%%%%% Data availability %%%%%%%%%%%%%%%%%%%%%%%%%%%%%%%%%%%%%%%%%%%%%%
% \section*{Data availability}

\printbibliography

% \clearpage
% \renewcommand\thefigure{S\arabic{figure}}
% \setcounter{figure}{0}    
% \section*{Supplementary Information}
% \subsection*{Supplementary Movies}
% Supplementary ...\\

\pagebreak

%%%%%%%%%%%%%%%%%%%%%%%%%%%%%%% Supplementary Material %%%%%%%%%%%%%%%%%%%%%%%%%%%%%%%%%%%%%%%%%%%%%%
\pagebreak
% \section*{Supplementary Material}

% \subsection*{A. Derivation of $\tau_p\to \infty$ entropic landscape}
%%%%%%%%%%%% Potentially delete, not necessary? vv %%%%%%%%%%%%%%%%%%%

\end{document}